\documentclass[aps,twocolumn,amsmath,amssymb,showpacs,prl,floatfix]{revtex4}
\usepackage{amsmath}
\usepackage{graphicx}
\usepackage{epsfig}
\newlength{\upit}\upit=0.1truein

\newcommand{\ltappr}{{{\lower4pthbox{$<$} } \atop \widetilde{ \ \ \ }}}
\newlength{\bxwidth}\bxwidth=1.5 truein

\newcommand{\dg}{^{\dagger }}

\newcommand \bea {\begin{eqnarray} }
\newcommand \eea {\end{eqnarray}}
\newcommand{\bk}{{\bf{k}}}

\newcommand{\bR}{{\bf{R}}}

\newlength{\figwidth}
\newlength{\shift}
\newcommand{\fg}[3]
{
\begin{figure}[ht]

\vspace*{-0cm}
\[
\includegraphics[width=\figwidth]{#1}
\]
\vskip -0.2cm
\caption{\label{#2}
\small#3
}
\end{figure}}

\begin{document}

\title{Tandem Pairing in Heavy Fermion Superconductors
}

\author{Rebecca Flint and Piers Coleman}
\affiliation{
Center for Materials Theory,
Rutgers University, Piscataway, NJ 08855, U.S.A. 
} 
\pacs{71.27.+a,74.20.Mn,74.25.Dw}

\begin{abstract}
We examine the internal structure of the heavy
fermion condensate, showing that it necessarily involves
a d-wave pair of quasiparticles on neighboring lattice sites, condensed in
tandem with a composite pair of electrons bound to a local moment,
within a single unit cell.  These two components draw upon the
antiferromagnetic and Kondo interactions to cooperatively enhance
the superconducting transition temperature.  The tandem condensate is electrostatically active, with a small electric
quadrupole moment coupling to strain that is predicted to lead to a superconducting shift in the NQR frequency.
\end{abstract}

\maketitle

In many strongly interacting materials, quasiparticles
are ill-formed at the superconducting transition, giving the Cooper pair a
non-trivial internal structure.  The 115 family of heavy fermion
superconductors\cite{hegger,monthoux,pagliusoPD,sarrao} provide an extreme
example of this phenomenon, where quasiparticle formation, through the screening
of local moments by electrons, coincides with the onset of
superconductivity.  

The 115 family has long attracted great interest for the remarkable
rise of the superconducting transition temperature from T$_{c}$=0.2K
in CeIn$_{3}$ under pressure\cite{mathur} to 2.3K in
CeCoIn$_5$\cite{hegger,monthoux,pagliusoPD} and then up to
18.5K in PuCoGa$_{5}$\cite{sarrao}.  
While the abundance of magnetism in the
phase diagram has led to a consensus that spin fluctuations drive the
superconductivity in the cerium compounds
\cite{miyake,scalapino,bealmonod,mathur,monthoux01}, the presence of unquenched local moments
at $T_c$ is difficult to explain within this
picture.  In a typical spin-fluctuation mediated heavy fermion
superconductor, the local moments quench to form a {\sl Pauli
paramagnet} ($\chi (T)\sim \chi_0$) well before the development of
superconductivity.  Yet NpPd$_5$Al$_2$\cite{aoki} and
Ce\{Co,Ir\}In$_5$\cite{monthoux,shishido} exhibit a Curie-Weiss
susceptibility, $\chi (T)\sim 1/ (T+T_{CW})$ down to $T_{c}$.   
Moreover, the highest transition temperatures are found in the actinide 115s, which show no signs of magnetism.

These observations led us to recently propose\cite{nphysus} that the
actinide 115s are \emph{composite pair
superconductors}\cite{elihu_wavefunction}, where the heavy Cooper pair
forms by combining two electrons in two orthogonal Kondo channels with
a spin flip to form a composite pair,
$\Lambda_C = \langle N| c\dg _{1\downarrow}c\dg_{2\downarrow}S_+|N+2\rangle$,
where $c\dg_{1,2}$ create electrons in two orthogonal
Kondo screening channels
\cite{CATK,nphysus}.  However, composite pairing
alone cannot account for the importance of magnetism in the Ce 115 phase
diagram.

We are led by these conflicting observations to propose a model for
the 115 materials where the composite and magnetic mechanisms work in
tandem to drive superconductivity.  Composite pairing originates from
two channel Kondo impurities, while magnetic pairing emerges from
antiferromagnetically coupled Kondo impurities.  These two systems are
equivalent at criticality in the dilute limit\cite{gan}, and we argue
that this connection persists to the lattice superconducting state that
conceals a common quantum critical point (QCP)\cite{Cox1996}.

\fg{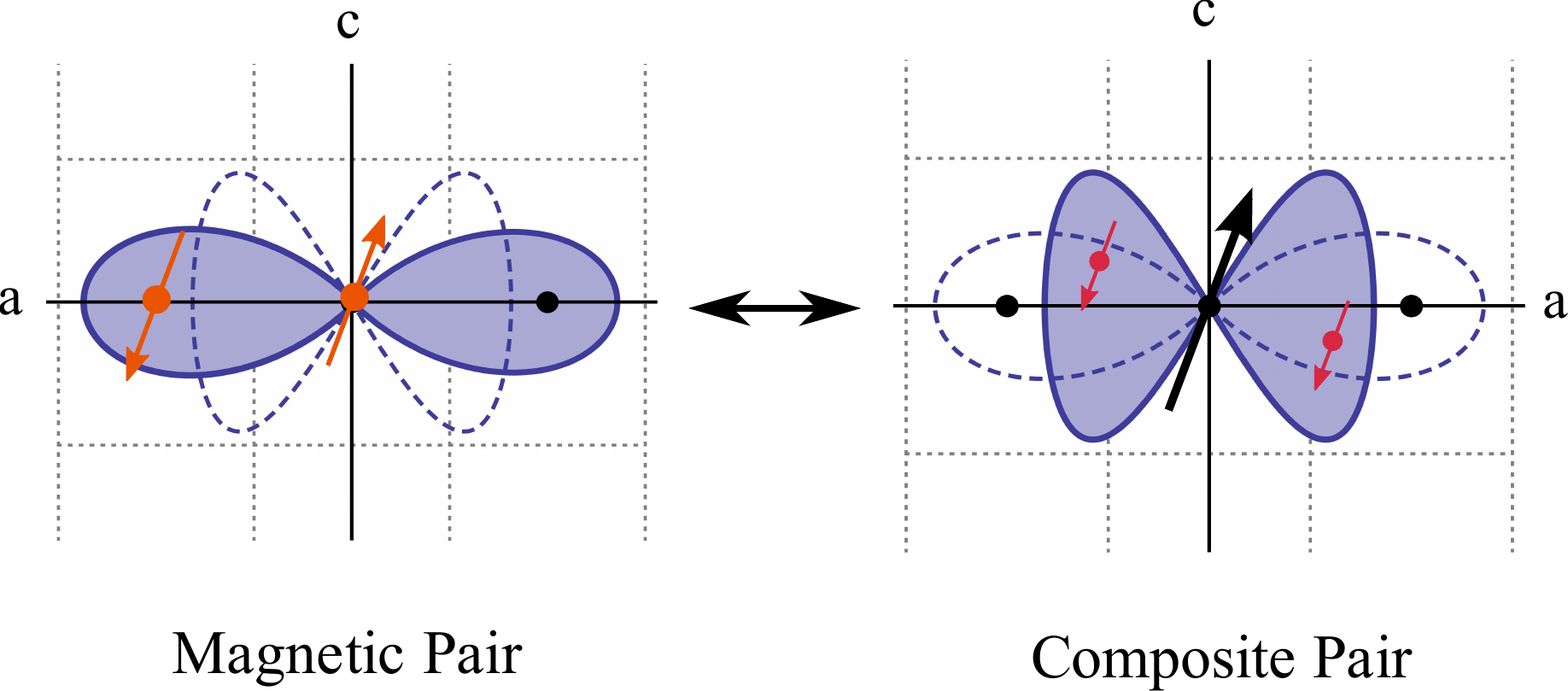}{pairs}{(Color online)A tandem pair contains a superposition of magnetic and composite pairing, both with d-wave symmetry.  The magnetic pair (left) contains neighboring f-electrons, while the composite pair (right) combines a spin flip and two conduction electrons.  The unit cell is denoted by dotted lines, with dots indicating the local moment sites.}

To expose the interplay between magnetic and composite pairing, we
examine the internal structure of a heavy fermion pair. 
In a Kondo lattice, the heavy quasiparticles are a linear combination 
$a\dg_{\bk \uparrow} = u_\bk c\dg_{\bk \uparrow} + v_\bk f\dg_{\bk \uparrow}$,
where $c\dg$ and $f\dg$ create conduction and localized electrons, respectively\cite{heavyQP}.   
The wavefunction is
\begin{equation}\label{}
|\Psi\rangle = P_{G}\exp(\Lambda\dg)|0\rangle,
\end{equation}
where $\Lambda\dg= \sum_{\bk}\Delta_{k}
(a\dg_{\bk \uparrow}a\dg_{-\bk \downarrow})$  creates a d-wave pair of
quasiparticles and $P_{G}$ is the Gutzwiller projection operator
restricting the number of f-electrons to one.
Acting the Gutzwiller projector on the f-electron reveals
its internal structure as a composite between a conduction electron
and a spin flip at a given site $j$, 
$P_{G}f_{j \uparrow}\dg \sim \bigl ( c_{j \downarrow}\dg S_+\bigr)P_{G}$.
The pairing field $\Lambda\dg $ contains three terms
\begin{equation}\label{l}
\Lambda\dg = \sum_\bk  \begin{pmatrix}
c\dg_{\bk \uparrow } &,f\dg_{\bk \uparrow}\end{pmatrix}
\left[\begin{matrix} \Delta^{e}_{\bk }
& \Delta^{C}_{\bk }\cr \Delta^{C}_{\bk } & \Delta^{M}_{\bk }
\end{matrix}\right]\left(\begin{matrix}
c\dg_{-\bk \downarrow}\cr f\dg_{-\bk \downarrow} \end{matrix}\right)
= \Psi_{e}\dg  
+ \Psi_{C}\dg+\Psi_{M}\dg.
\end{equation}
The diagonal terms, with $\Delta^{e}_{\bk }= u_{\bk }^{2}\Delta_{\bk}$ 
and $\Delta^{M}_{\bk }= v_{\bk }^{2}\Delta_{\bk}$
create f- and conduction electron pairs.  A d-wave
pair of f-electrons is an inter-site operator, taking
the form
\begin{equation}
\Psi\dg _{M} = \sum_{i,j}\Delta^{M} ({\bf{R}}_{ij})\biggl[(c\dg_{i \uparrow} S_{i-})(c\dg_{j \downarrow} S_{j +})\biggr]
\end{equation}
outside the Gutzwiller projection.
However, if we expand the 
off-diagonal terms in real space,
\begin{equation}
\Psi\dg _{C} = \sum_{i,j}\Delta^{C} ({\bf{R}}_{ij}) \biggl[c\dg_{i \uparrow} c\dg_{j \uparrow}S_{j -}\biggr]
\end{equation}
where $\Delta^{C} ({\bf R}) = \sum_{\bk } (u_{\bk}v_{\bk}\Delta_{\bk})e^{i \bk
\cdot {\bf R}}$,
we find a composite pair formed between 
a triplet pair of conduction electrons and a spin
flip\cite{nphysus,CATK,elihu_wavefunction}. 
Unlike its diagonal counterparts, which are necessarily inter-site,
composite pairs are compact objects formed from
pairs of orthogonal Wannier states surrounding
a {\sl single local moment} (Fig. \ref{pairs}).

Magnetic interactions favor the 
inter-site  component of the pairing, while the two-channel
Kondo effect favors the composite intra-site component. 
However, both components will always be present in the superconducting Kondo lattice.
If the product of the Kondo screening channels has a
d-wave symmetry, the composite and magnetic order parameters necessarily
couple linearly to one another, a process that enhances the transition temperature over a large
region of the phase diagram, providing a natural explanation for both the actinide and Ce 115s.

To treat these two pairing mechanisms simultaneously, we introduce the
two channel Kondo-Heisenberg model, 
\begin{equation}
H = H_c + H_{K1} + H_{K2} + H_{M}
\end{equation}
and solve it in the symplectic-$N$ limit\cite{nphysus}, where
\bea
H_c & = & \sum_{\bk} \epsilon_{\bk} c_{\bk\sigma}\dg c_{\bk\sigma},\quad
H_{M} = J_{H}\sum_{\langle ij\rangle} \vec{S}_i \cdot \vec{S}_j\\
H_{K\Gamma} & = & J_{\Gamma}\sum_{j}\psi\dg _{j\Gamma a}\vec{\sigma}_{ab}\psi_{j\Gamma b}\cdot \vec{S}_j.
\eea
where $\vec{S}_j$ is the local moment on site $j$, and $\psi_{j\Gamma}$ is the Wannier state representing a conduction electron on site $j$ with symmetry $\Gamma$,
\begin{equation}
\psi_{j\Gamma a} = \sum_\bk  \Phi_{\Gamma \bk ab}c_{\bk b} \mathrm{e}^{i\bk\cdot \bR_j},
\end{equation}
where the form factor $\Phi_{\Gamma k ab}$ is only diagonal in the spin indices  in the absence of spin-orbit.  
Microscopically, the two 
orthogonal Kondo channels, $J_\Gamma$
arise from virtual fluctuations from
the ground state doublet to excited singlets, where the two channels
correspond to adding and removing an electron, respectively.  The Ce $4f^1$ state is split by tetragonal symmetry into three Kramer's doublets, where
$\Gamma_7^+$ is the ground state doublet\cite{ChristiansonCF,nakatsujiCF}, so we may summarize the virtual valence fluctuations with:
\begin{equation}
4f^0 (\cdot)\overset{\Gamma_7^+}{\rightleftharpoons} 4f^1 \left(\Gamma_7^+\right) \overset{\Gamma_6}{\rightleftharpoons} 4f^2 \left(\Gamma_7^+ \otimes \Gamma_6\right).
\end{equation}
Requiring the composite pairing to resonate with the d-wave magnetic pairing\cite{greene} uniquely selects $\Gamma_7^+ \otimes \Gamma_6$ as the lowest doubly occupied state, as this combination leads to d-wave composite pairing\cite{nphysus}.  
A simplified two dimensional model is sufficient to illustrate the basic physics, where the d-wave composite pair now comes from the combination of s-wave hybridization in channel one and d-wave hybridization in channel two\cite{ghaemi,weber}.
The magnetism is included as
an explicit RKKY interaction, $J_H$ between
neighboring local moments $\langle ij \rangle$, generated by integrating out electron in bands
far from the Fermi surface.  Treating the magnetism as a Heisenberg term leads to a 
two band version of resonating valence bond (RVB) superconductivity\cite{RVB}, where the local moments form
valence bonds which ``escape'' into the conduction sea through the Kondo hybridization to form charged, mobile Cooper pairs\cite{andrei}.

To solve this model, we use a fermionic spin representation,  $\vec{S}_j = f\dg_j \vec{\sigma}_N f_j$; symplectic-$N$ maintains the time-reversal properties of $SU(2)$ in the large $N$ limit by using the symplectic generalization of the Pauli matrices $\vec{\sigma}_N$ to construct the spin Hamiltonians\cite{nphysus},
\bea
H_{K\Gamma}(j) & = & -\frac{J_{\Gamma}}{N}\left[(\psi\dg _{j\Gamma}f_{j}) 
(f\dg_j\psi_{j\Gamma})+(\psi \dg_{j\Gamma} \epsilon\dg f\dg _j)(f_{j} 
\epsilon \psi _{j\Gamma})
\right]\nonumber\\
H_{M}(ij) & = & -\frac{J_{H}}{N}\left[(f\dg _{i}f_{j}) 
(f\dg_j f_{i})+(f \dg_{i} \epsilon\dg f\dg _j)(f_{j} 
\epsilon f _{i}),
\right]
\eea
where $\epsilon$ is the large $N$ generalization of $i \sigma_2$.  Each quartic term can be decoupled by a Hubbard-Stratonovich field, leading to normal, $V_{\Gamma}$ and anomalous, $\Delta_\Gamma$ hybridization in each Kondo channel and particle-hole, $h_{ij}$ and pairing, $\Delta_{ij}^H$ terms for the spin liquid.  The $SU(2)$ gauge symmetry of the Hamiltonian, $f \longrightarrow uf+ v\epsilon\dg f\dg$ is used to eliminate $\Delta_1$.  The lowest energy solutions contain only pairing fields in the magnetic and second Kondo channels, giving rise to three Hubbard-Stratonovich fields, $V_1$, $\Delta_2$ and $\Delta_H$, where $\Delta_H$ is d-wave in the plane, so that $\Delta_{k}^H \equiv \Delta_H (\cos k_x - \cos k_y)$.  Using the Nambu notation,
$\tilde{c}\dg _{\bk }= (c\dg_{\bk},\epsilon c_{-\bk})$, 
$\tilde{f}\dg _{\bk }= (f\dg_{\bk},\epsilon f_{-\bk})$,
and defining
$\mathcal{V}_k = V_{1} \Phi_{1 \bk}+\Delta_{2}
\Phi_{2 \bk}$, the mean field Hamiltonian can be concisely written
as
\bea
\label{fullH}
H &=& \sum_k 
\left(
\begin{array}{cc}
\tilde{c}_{k}\dg & \tilde{f}_{k}\dg
\end{array}
\right)
\left[
\begin{array}{cc}
\epsilon_k \tau_3 & \mathcal{V}_k\dg \\
\mathcal{V}_k & \lambda \tau_3 + \Delta_{Hk}\tau_1
\end{array}
\right]
\left(
\begin{array}{c}
\tilde{c}_{k} \\
\tilde{f}_{k}
\end{array}
\right)\nonumber\\
&&+ N\left(\frac{V_{1}\dg V_{1}}{J_1}+\frac{\Delta_{2}\dg \Delta_{2}}{J_2}+\frac{4\Delta_H^2}{J_H}\right),
\eea
where $\lambda$ is the Lagrange multiplier enforcing the constraint $n_f = 1$.  The mean field Hamiltonian can be diagonalized analytically.  Upon minimizing the free energy, we obtain four equations for $\lambda, V_1, \Delta_2,$ and $\Delta_H$.  Solving these numerically, and searching the full parameter space of $J_2/J_1$, $J_H/J_1$ and $T$ to find both first and second order phase transitions, we find four distinct phases: a light Fermi liquid with free local moments when all parameters are zero, at high temperatures; a heavy Fermi liquid when either $V_1$ or $\Delta_2$ are finite, with symmetry $\Gamma$, below $T_{K\Gamma}$; a spin liquid state decoupled from a light Fermi liquid when $\Delta_H$ is finite, below $T_{SL}$; and a tandem superconducting ground state with $V_1$, $\Delta_2$ and $\Delta_H$ all finite, below $T_c$, as shown in Fig. \ref{phase}.  There is no long range magnetic order due to our fermionic spin representation.

\fg{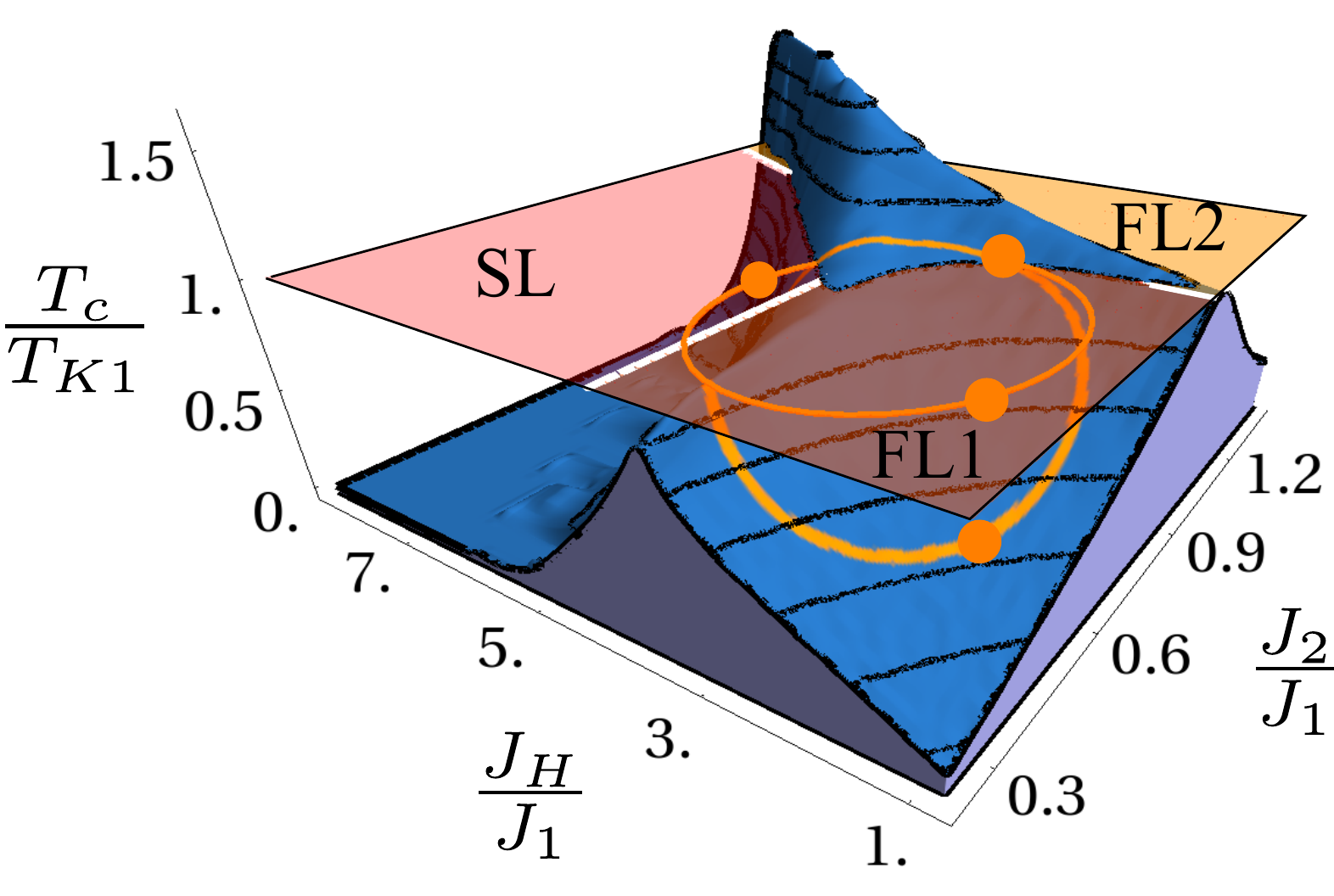}{phase}{(Color online) The superconducting transition
temperature as the amounts of magnetic, $J_H$ and second channel, $J_2$ couplings are varied.   A slice at $T = T_{K1}$ shows the regions of the spin liquid and Fermi liquids, and the orange ellipse is a path illustrating how materials could tune the relative coupling strengths (see Fig. \ref{path}).  The phase diagram was calculated in a simple two dimensional model with channel one s-wave and channel two d-wave ($n_c = .75$).   The transition is first order for $J_H/J_1 > 4$, but otherwise second order.}

Experimentally, Ce$M$In$_5$ can be continuously tuned from $M =$ Co to Rh to Ir\cite{pagliusoPD}. 
While CeRhIn$_5$ is a canonical example of a magnetically paired superconductor, where moderate pressure reveals a superconducting dome as the N\'{e}el temperature vanishes\cite{hegger}, further pressure\cite{muramatsu} or Ir doping on the Rh site\cite{pagliusoPD} leads to a second dome, where spin fluctuations are weaker\cite{nqrIrRh}. 
We assume that the changing chemical pressure varies the relative strengths of the Kondo and RKKY couplings, so that doping traces out a path through the phase diagram like the one in Fig. \ref{path}, chosen for its similarities to Ce$M$In$_5$.  While different paths may lead to one, two or three superconducting domes, by maintaining the same Fermi liquid symmetry throughout ($T_{K1} > T_{K2}$), we are restricted to one (magnetic only) or two (magnetic and tandem) domes.

\fg{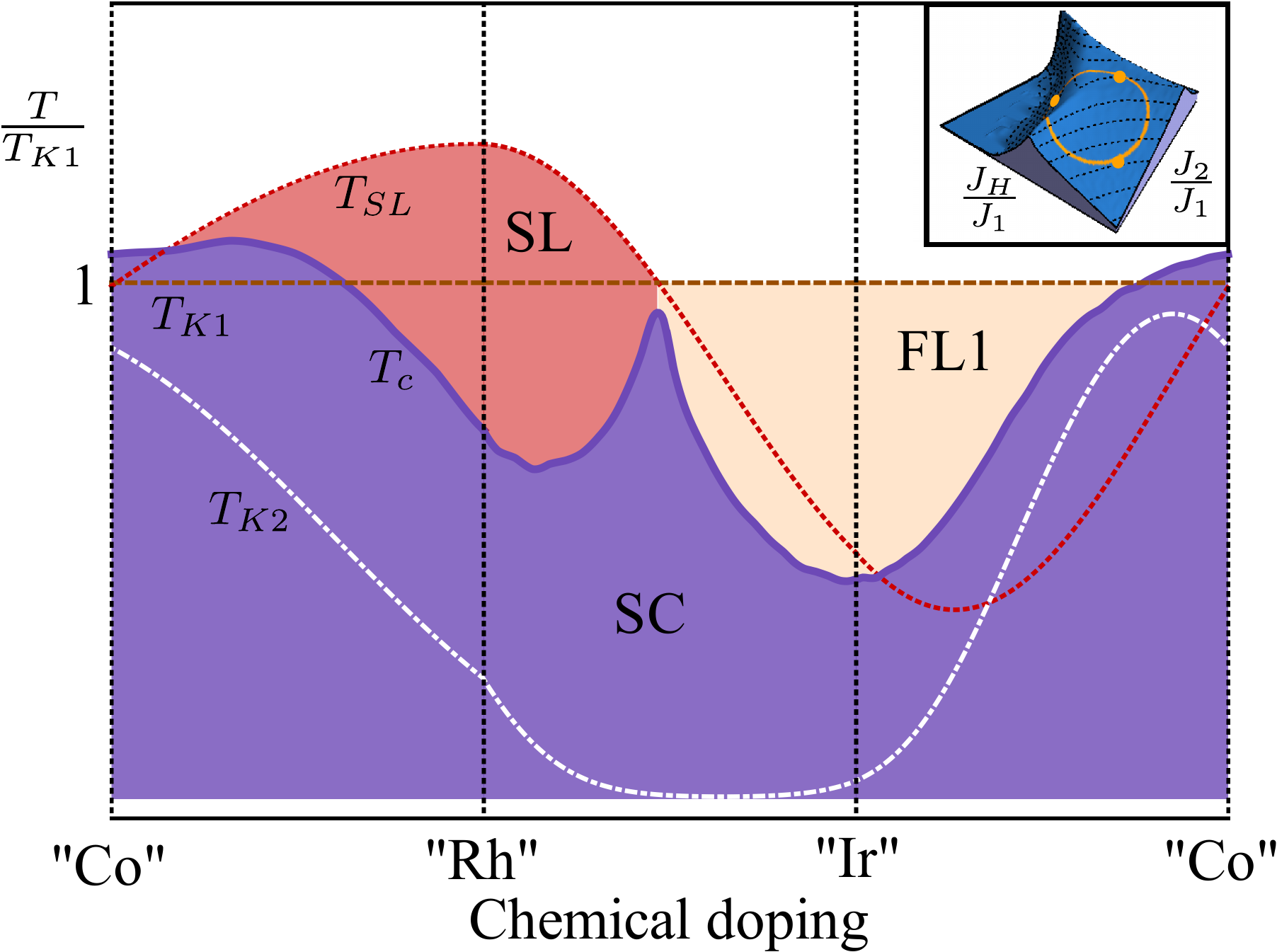}{path}{(Color online) A possible experimental path through the phase diagram in Fig \ref{phase}, chosen for its similarity to the Ce 115 doping phase diagram\cite{pagliusoPD}.  The transition temperatures for superconductivity ($T_c$ in solid blue), spin liquid ($T_{SL}$ in dotted red), and Fermi liquid ($T_{K1}$ in dashed orange and $T_{K2}$ in dot-dashed white) are plotted for comparison.  Temperatures are scaled by $T_{K1}$, which may itself vary as one moves around the phase diagram\cite{willers}.  While we always find a superconducting ground state, due to our choice of a fermionic spin representation, real materials will have an antiferromagnetic ground state for $T_{SL}/T_{K1}$ sufficiently large.}

A qualitative understanding of this tandem pairing can be obtained within a
simple Landau expansion. For $T \sim T_c \ll T_{K1}$, 
$\Phi \equiv \Delta_{2}$ and $\Psi \equiv \Delta_H$ will be small, and the free energy can be expressed as
\bea
\label{landau}
F & = & \alpha_1(T_{c1}-T) \Psi^2 + \alpha_2(T_{c2}-T) \Phi^2 + 2 \gamma \Psi \Phi \nonumber\\
& + & \beta_1 \Psi^4 + \beta_2 \Phi^4 + 2 \beta_i \Psi^2 \Phi^2
\eea
$\alpha_{1,2}$, $\beta_{1,2,i}$ and $\gamma$ are all functions of $\lambda$ and $V_1$ and can be calculated exactly in the mean field limit.  The linear coupling of the two order parameters, $\gamma = \partial F/\partial \Delta_{2} \partial \Delta_H$ is always nonzero in the heavy Fermi liquid, leading to an enhancement of the transition temperature,
\begin{equation}
T_c = \frac{T_{c1} + T_{c2}}{2} + \sqrt{\left(\frac{T_{c1}-T_{c2}}{2}\right)^2+\frac{\gamma^2}{\alpha_1 \alpha_2}}.
\end{equation}
For $\beta_1 \beta_2 > \beta_i^2$, the two order parameters are only weakly repulsive, leading to smooth crossovers from magnetic to composite pairing under the superconducting dome\cite{mineev}.

While the development of conventional superconductivity does not change the underlying charge
distribution, tandem pairing is electrostatically
active, as composite pairing redistributes charge, leading to an
electric quadrupole moment.  The transition temperature of the 115
superconductors is known to increase linearly with the lattice $c/a$
ratio\cite{bauer}, conventionally attributed to decreasing
dimensionality.  Our theory suggests an alternative interpretation: in
a condensate with a quadrupole moment, $Q_{zz} \propto \Psi_C^2$,
which couples linearly to the tetragonal strain, $\Delta F
\propto-Q_{zz} u_{tet}$, the second term in the Landau
free energy (\ref{landau}) becomes $\alpha_2 [T - (T_{c2}+\lambda u_{tet})]
\Psi_C^2$, naturally accounting for the linear increase
in $T_c$.  
The development of a condensate quadrupole moment should be
also detectable as a shift of the nuclear quadrupole resonance (NQR)
frequency at the nuclei of the surrounding ions.

The link between f-electron valence and the Kondo effect is well
established\cite{gunnarsson}, but 
tandem pairing introduces a new element to this relationship.
Changes in the charge distribution around the Kondo ion can be read off
from its coupling to the changes in the chemical potential, $\Delta
\rho(x) = |e|\delta H/\delta \mu(x)$.  The sensitivity of the
Kondo couplings to $\mu$ is obtained from a
Schrieffer-Wolff transformation of a two-channel Anderson model, which
gives $J_\Gamma^{-1} = \Delta E_{\Gamma}
/V_{\Gamma,0}^2$.  Here, $V_{\Gamma,0}$ are the bare hybridizations and $\Delta E_\Gamma$ are the charge excitation
energies.  With a shift in $\mu \rightarrow \mu
+ \delta \mu(x)$, $\delta J_\Gamma^{-1} =
\pm \vert \Phi_{\Gamma} (x)\vert^{2} \delta\mu(x)/V_{\Gamma,0}^{2}$.
The sign is positive for $J_1$ and negative for $J_2$ because they involve
fluctuations to the empty and doubly occupied states, respectively:
$f^0 \overset{\Gamma_1}{\rightleftharpoons} f^1
\overset{\Gamma_2}{\rightleftharpoons} f^2$.  Differentiating
(\ref{fullH}) with respect to $\delta \mu (x)$, the
change in $\rho(x)$ will be:
\begin{equation}
\label{charge_density}
\Delta \rho(x)= |e| \left[ 
\left(\frac{V_1}{V_{1,0}} \right)^{2}
|\Phi_{1}(x)|^2 - 
\left(\frac{\Delta _2}{V_{2,0}} \right)^{2}
|\Phi_{2}(x)|^2 
\right].
\end{equation}
For equal channel strengths, the total charge is
constant, and the f-ion will develop equal hole densities in $\Gamma_7^+$ and
electron densities in $\Gamma_6$, leading to a positive change in the
electric field gradient, $\partial E_z/\partial z \propto (T_{c}-T) > 0$ 
at the in-plane In site that will appear as a shift in the NQR frequencies growing abruptly below $T_{c}$ (see Figure \ref{charge}). 

\fg{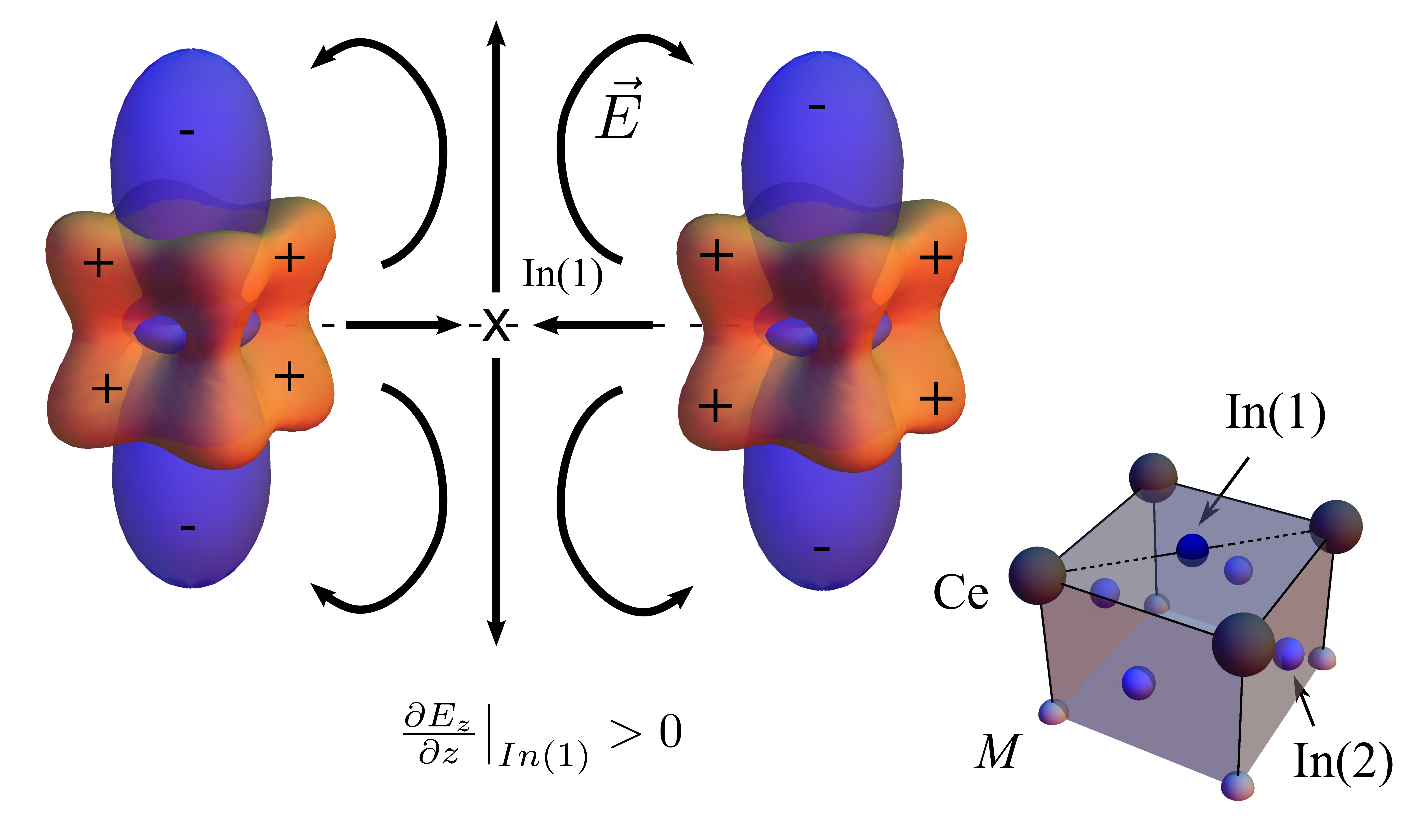}{charge}{(Color online) Predicted NQR frequency shift, $\Delta\nu_{NQR}$ in Ce$M$In$_5$.  The inset shows the relative locations of the indiums in-, In(1) and out-of-plane, In(2).  $\Delta\nu_{NQR}$ measures
the change in the electric field gradient (EFG) due to the onset of superconductivity.  
For equal channel strengths,
the total charge of the f-ion remains unity, but the increasing
occupations of the empty and doubly occupied sites cause holes to
build up with symmetry $\Gamma_7^+$ (orange) and electrons with symmetry
$\Gamma_6$ (blue).  The change in charge distribution and resulting electric fields are shown above in a slice along the [110] direction (dashed line in the inset).  The positive EFG, $\partial E_z/\partial z$ at the In(1) site will lead to a sharp positive shift in $\nu_{NQR}$ starting at $T_c$. }

The f-electron valence should also contain a small superconducting shift, observable with core-level X-ray spectroscopy,
obtained by integrating
(\ref{charge_density}): $\Delta n_f(T) \propto \Psi_C^2 \propto (T_{c}-T)$, as $\Psi_C \propto \Delta_2$ when $J_1 > J_2$.  
While the development of Kondo screening leads to a gradual valence decrease 
through $T_K$, as it is a crossover scale,
the development of superconductivity is a phase transition, leading to
a sharp mean-field increase. Observation of sharp shifts at $T_c$ in either the NQR
frequency or the valence would constitute an unambiguous confirmation
of the electrostatically active tandem condensate.

The authors would like to thank S. Burdin, C. Capan, Z. Fisk, H. Weber, R. Urbano, and particularly M. Dzero for discussions related to this work.  This research was supported by National Science Foundation Grant DMR-0907179.
%
%
%
%

\end{document}